# The Physics of Antimicrobial Activity of Ionic Liquids

V. K. Sharma[1,2*], J. Gupta[1,2], J. Bhatt Mitra[3], H. Srinivasan[1,2], V. García Sakai[4], S. K. Ghosh[5] and S. Mitra[1,2]

[1]Solid State Physics Division, Bhabha Atomic Research Centre, Mumbai, 400085, India

[2]Homi Bhabha National Institute, Mumbai, 400094, India

[3]Radiopharmaceuticals Division, Bhabha Atomic Research Centre, Mumbai 400085, India

[4] ISIS Neutron and Muon Source, Science and Technology Facilities Council, Rutherford Appleton Laboratory, Didcot OX11 0QX, United Kingdom

[5]Department of Physics, School of Natural Sciences, Shiv Nadar Institution of Eminence, NH91, Tehsil Dadri, G. B. Nagar  Uttar Pradesh 201314, India

*Corresponding Author: Email: sharmavk@barc.gov.in; vksphy@gmail.com Phone +91-22-25594604

# Abstract

The bactericidal potency of ionic liquids (ILs) is well-established, yet their precise mechanism of action remains elusive. Here, we show evidence that the bactericidal action of ILs primarily involves permeabilizing the bacterial cell membrane. Our findings reveal that ILs exert their effects by directly interacting with the lipid bilayer and enhancing the membrane dynamics. Lateral lipid diffusion is accelerated which in turn augments membrane permeability, ultimately leading to bacterial death. Furthermore, our results establish a significant connection: an increase in the alkyl chain length of ILs correlates with a notable enhancement in both lipid lateral diffusion and antimicrobial potency. This underscores a compelling correlation between membrane dynamics and antimicrobial effectiveness, providing valuable insights for the rational design and optimization of IL-based antimicrobial agents in healthcare applications.

**Graphical Abstract**

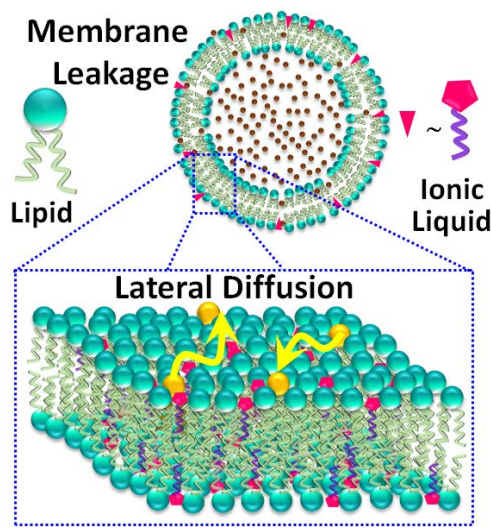

The emergence of antimicrobial resistance poses a significant threat to public health, fueling the urgent quest for alternative therapeutic strategies[1-2]. According to data from the World Health Organization (WHO), antibiotic-resistant bacterial diseases contribute to the deaths of at least 700,000 individuals annually[2]. WHO cautions that without decisive action against antibiotic-resistant bacteria, the annual death toll could surge to 10 million people by the year 2050. In the ongoing battle against bacterial resistance to antibiotics, there is a crucial need for novel alternatives that can better withstand evolving resistance mechanisms. Ionic liquids (ILs) have emerged as promising candidates in this pursuit, drawing attention due to their distinctive physicochemical properties and versatile biological activities [3-6]. ILs, characterized by their unique ionic nature and low melting points, have demonstrated remarkable antimicrobial efficacy against a diverse array of pathogens, including bacteria, fungi, and viruses[3-5, 7-11]. Their antimicrobial properties have offered a glimpse of hope in the ongoing battle against antibiotic-resistant pathogens, opening new avenues for therapeutic intervention. A key advantage of ILs lies in their highly customizable structure, allowing for precise tailoring of their chemical, physical, and biological properties[12-13]. This versatility makes ILs promising candidates for combating bacterial infections, offering opportunities for targeted antimicrobial strategies. Despite the considerable evidence regarding the antimicrobial activity of ILs, their fundamental mechanism of action remains poorly understood. While it is widely accepted that ILs exert bactericidal effects primarily by interacting with bacterial cell membranes[10, 14-18], the specific molecular mechanisms underlying this process continue to be a subject of ongoing research. ILs may eliminate bacteria through various mechanisms, including the disruption of microbial cell membranes, acting on intracellular components of microorganisms, inhibition of essential cellular processes, and interference with metabolic pathways. Deciphering the precise mechanism by which ILs potentially kill bacteria will provide a holistic perspective on their utility and aid in tailoring the ILs to target different microbial species effectively.

In this study, we unveil the fundamental mechanism underlying the antimicrobial action of imidazolium-based ILs, as the preferred ILs by the pharmaceutical industry due to the prevalence of the N-substituted imidazolium ring in many natural products and bioactive molecules involved in human metabolism[19-20]. Using minimum inhibitory concentrations (MIC) and minimum bactericidal concentration (MBC) measurements, we have shown the antimicrobial potential of these ILs. Through the use of various imidazolium-based ILs with differing alkyl chain lengths and anions, we elucidate the role of IL hydrophobicity in antimicrobial activity, emphasizing that the direct interaction with the lipid membrane and enhanced membrane permeability, are primary contributors to their efficacy. Cell membrane permeability is associated with the membrane fluidity which is directly related to membrane dynamics. By using neutron scattering, we demonstrate that the dominant effect of the IL with the longer chain length on the lateral diffusion is correlated with its higher membrane permeability ultimately resulting in higher antimicrobial activity.

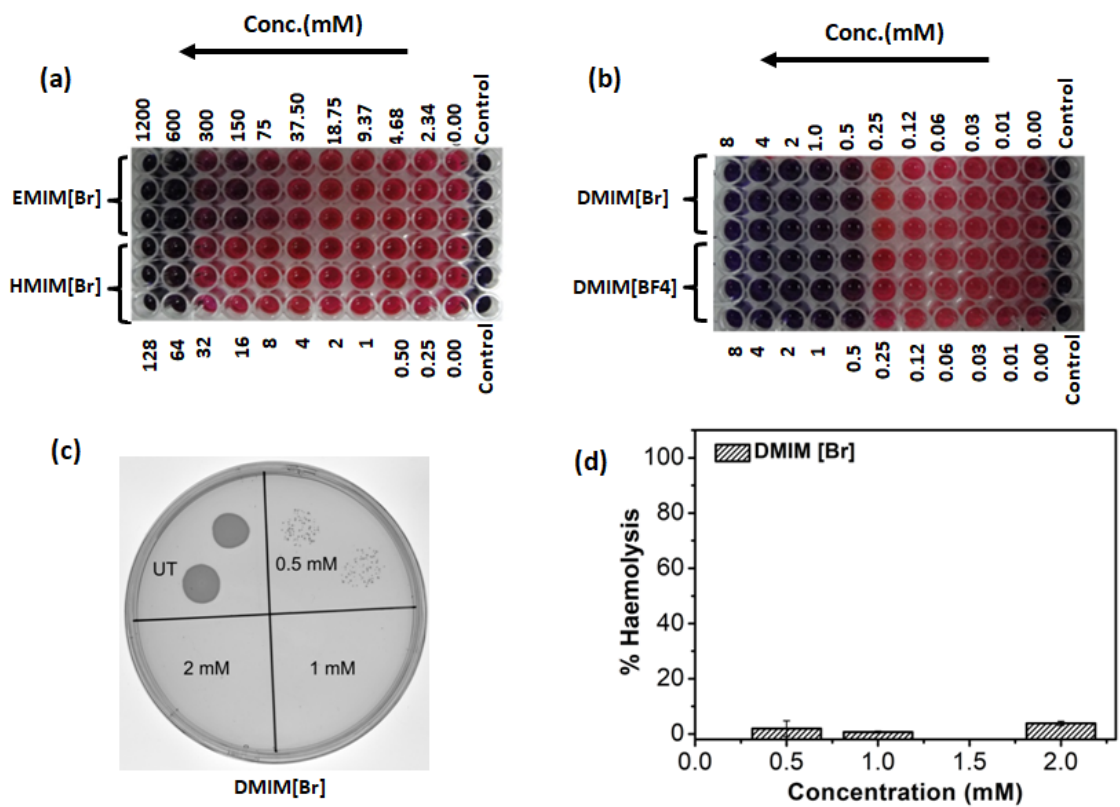

**Fig. 1** (a-b) A representative of resazurin assay performed for imidazolium based ILs having varied alkyl chain length cations and anions with *S. aureus* ATCC 25923. (c) Spot assay showing bactericidal action of DMIM[Br]. (d) Bar graph showing % haemolysis caused by DMIM[Br].

MIC for *S. aureus* ATCC 25923 cells were determined using a resazurin reduction assay in a microtiter format. Metabolically active cells reduce resazurin dye to resorufin. MIC for this assay was defined as the lowest concentration of compound at which a reduction of resazurin (blue) into resorufin (pink) stopped. Four different ILs namely 1-ethyl-3-methyl-imidazolium bromide (EMIM[Br] or C2MIM[Br]), 1-hexyl-3-methylimidazolium bromide (HMIM[Br] or C6MIM[Br]), 1-decyl-3-methylimidazolium bromide  (DMIM[Br] or C10MIM[Br]), and 1-decyl-3-methylimidazolium tetrafluoroborate (DMIM[BF4] or C10MIM[BF4]) having either different chain lengths or different anions were studied. All ILs inhibit the growth of *S. aureus* under our experimental conditions, as shown in Fig. 1 (a-b). MICs for each IL have been obtained. MIC values decreased significantly (by orders of magnitude) with increasing aliphatic chain length, suggesting that the alkyl chain length of the IL plays a very substantial role in the IL's bactericidal efficacy. The MIC values fall from 600mM to 64mM to 0.5mM in going from EMIM[Br] to HMIM[Br] to DMIM[Br]. In contrast, changing the anion does not lead to major changes (MIC for DMIM[BF4] is 0.5mM). To assess bactericidal activity, spot assays were conducted using DMIM[Br], the IL with the longest alkyl chain. Our findings indicate that DMIM[Br] exhibits bactericidal properties, with a MBC of 1 mM, which is only twice the value of the MIC, as illustrated in Fig. 1(c). In the pursuit of harnessing the therapeutic potential of ILs, a critical imperative lies in their ability to discriminate between bacterial and human cells. The effectiveness of ILs relies on their ability to selectively target bacterial cells while maintaining the viability of human cells.

Bacterial and mammalian cells membranes have very different chemical compositions. A key difference being the abundance of cholesterol (up to 40 mol%) in mammalian cell membranes, which is absent in their bacterial counterparts. This compositional difference between the two membrane types plays an important role in the selectivity of the ILs. For example, it has been shown that disrupting effects of ILs are considerably mitigated in the presence of cholesterol[17,21]. We have carried out hemolysis assays to assess the potential cytotoxicity of DMIM[Br] on human red blood cells (RBC) which plasma membrane has a significant amount of cholesterol (about 40 mol %). The obtained hemolysis data are presented in Fig. 1(d). Remarkably, this IL exhibits no toxicity towards RBC at concentrations up to 2 mM (twice the MBC), suggesting selective bactericidal activity particularly at lower concentrations. Therefore, DMIM[Br] shows promising antibacterial potential for therapeutic applications.

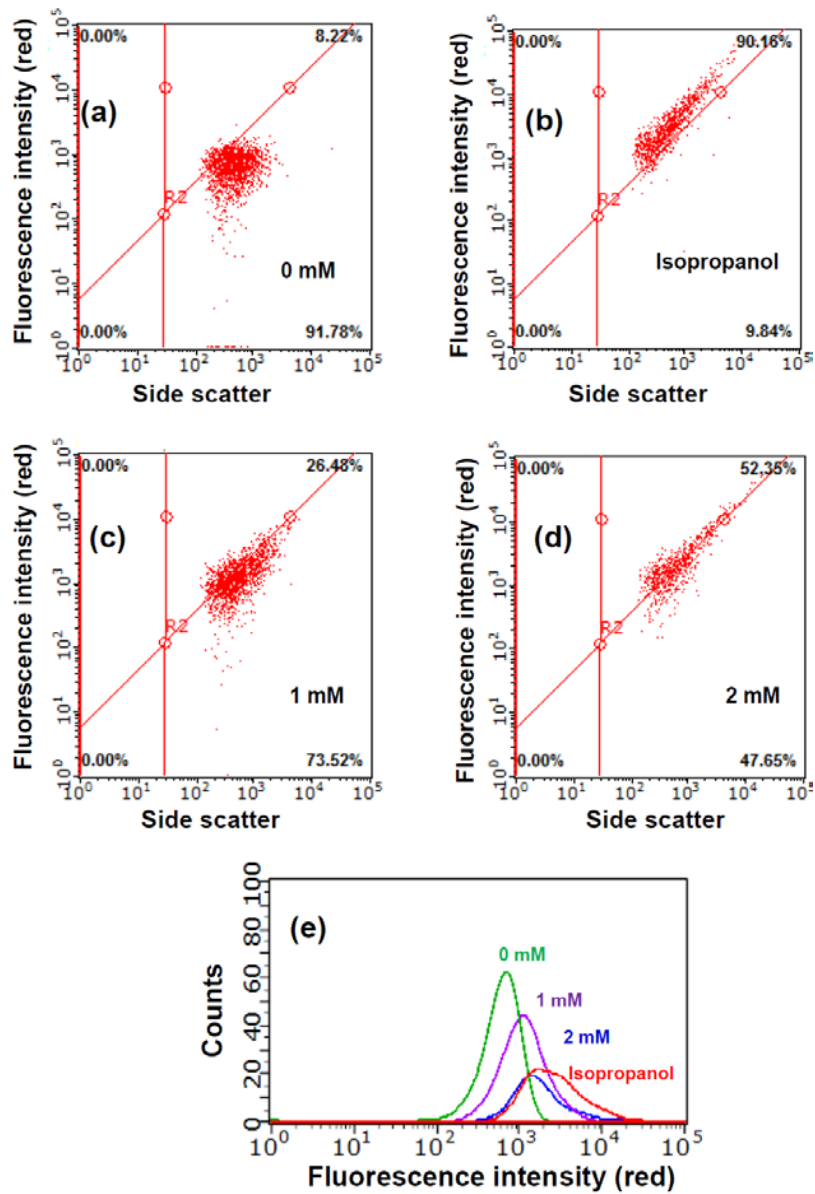

**Fig. 2** Representative red fluorescence (using PI) *vs* side scatter dot plots for *S. aureus* ATCC 25923 cells treated either with (a) 0 mM DMIM[Br], (b) 70% isopropanol  (c) 1mM (MBC) DMIM[Br], (d) 2mM (2×MBC) DMIM[Br] and (e) Bacterial cell count *vs* red fluorescence intensity plotted as a histogram.

To investigate the mechanism underlying the antibacterial effects of ILs, flow cytometry measurements[22] were conducted on S. *aureus* with different concentrations of DMIM[Br]. Propidium iodide (PI) red dye can only enter permeabilized bacterial cells and not live cells. This property of PI was utilized to distinguish between live (0 mM DMIM[Br]) and permeabilized bacterial populations (70% isopropanol treated bacteria) as shown in Fig. 2(a) and (b), respectively. These bacterial populations could be distinguished on red fluorescence and side scatter dot plots using region R2 (Fig. 2). As the concentration of DMIM[Br] was increased from 0 mM (Fig. 2 (a)) to MBC and 2 × MBC (Fig. 2 (c) and Fig. 2 (d)), the bacterial population started to fall in the permeabilized region much like the control treated with 70% isopropanol as shown in Fig. 2 (b). Therefore, the results indicate that the percentage of permeabilized bacteria increase with increasing concentration of DMIM[Br]. As shown in Fig. 2(e), the mean fluorescence intensity (MFI) of PI increased with increasing concentration of DMIM[Br]. The increased MFI is indicative of PI accumulation, which in turn suggests that DMIM[Br] effectively permeabilizes bacterial cells.

Though it is clear that DMIM[Br] exerts its bactericidal effect by disturbing the bacterial cell membrane, the underlying mechanism at a molecular level remains unclear. Whether DMIM[Br] directly interacts with the lipid membrane or targets specific receptors on the S. *aureus* membrane for disruption is yet to be determined. To address this, we examine how the presence of ILs affects the permeability of unilamellar vesicles (ULVs) of a model membrane, dipalmitoylphosphatidylcholine (DPPC), using a dye leakage assay. DPPC was selected due to its well-defined phase behavior and structural characteristics and has been prevalently used as a model bacterial membrane[23-25]. Although simple, this robust model system is used to understand the microscopic physics behind the action mechanism of ILs which can build a solid foundation for future studies that can be carried out on complex membrane systems akin to more biologically relevant compositions. Fig. 3 (a) illustrates the extent of dye leakage (%) as a function of DMIM[Br] concentration, measured 15 minutes post-addition. It is evident that DMIM[Br] permeabilizes the DPPC membrane; increasing the DMIM[Br] concentration leads to accelerated dye leakage kinetics which reach a maximum value at ~ 1mM DMIM[Br], coinciding with the MBC value of DMIM[Br]. This observation confirms that the bactericidal activity of IL is directly linked to its interaction with the lipid membrane. This positions ILs as highly effective antimicrobial agents capable of targeting a broad spectrum of pathogens. Dye leakage assays conducted on the shorter chain length IL, HMIM[Br], showed no notable dye leakage up to a concentration of 2mM, consistent with its MIC value (~ 64 mM). This highlights the significance of IL hydrophobicity in antimicrobial activity, a property which can be augmented by extending the alkyl chain length of ILs.

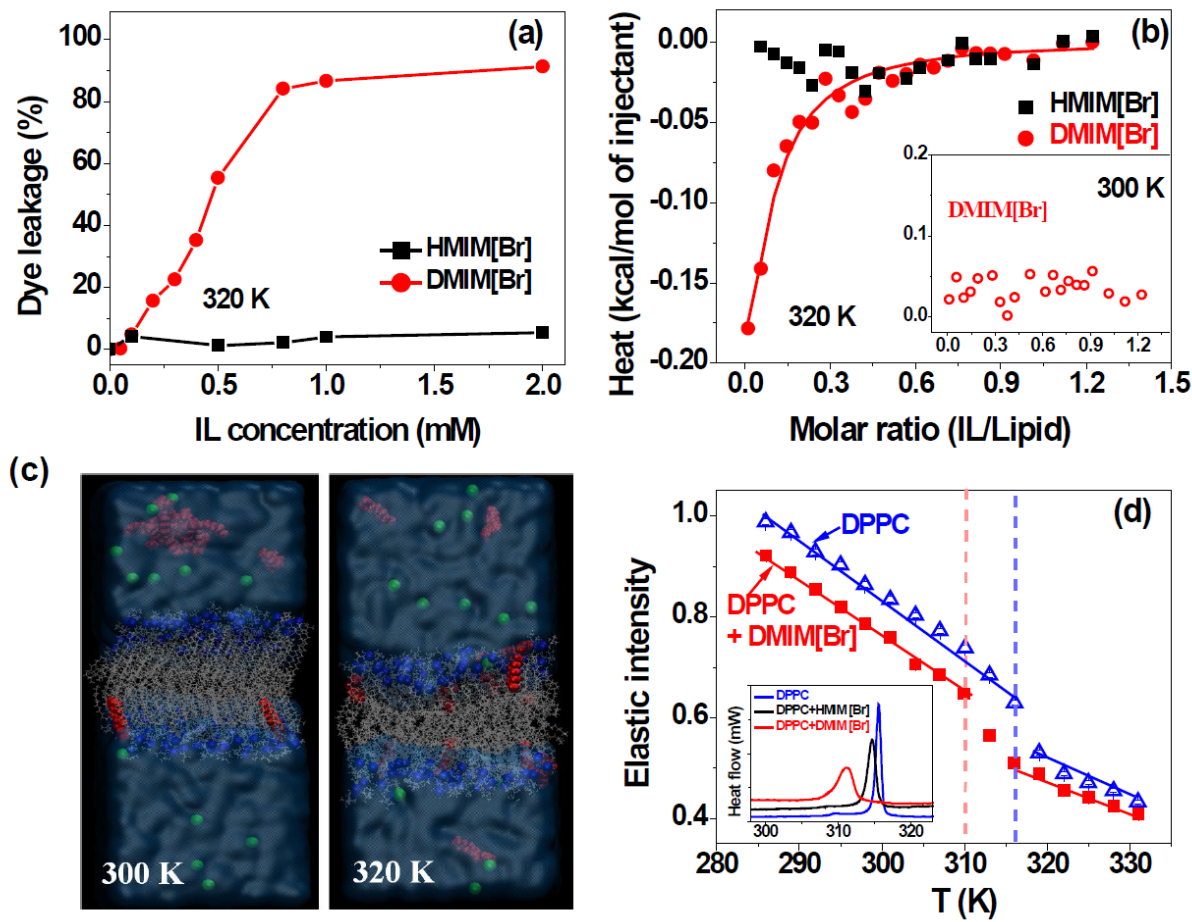

**Fig. 3** (a) The extent of calcein leakage from DPPC ULVs induced by HMIM[Br] and DMIM[Br] 15 minutes after addition. (b) Binding isotherms of HMIM[Br] and DMIM[Br] with DPPC in the fluid phase. The inset shows binding isotherms obtained for DMIM[Br] with DPPC in the gel phase (c) Snapshots of the DPPC+ DMIM[Br] simulation in the (i) gel (300 K) and (ii) fluid (320 K) phases. ILs are shown in red, Br are shown in cyan. (d) *Q*-averaged elastic intensity scans for DPPC ULVs with and without DMIM[Br] in the heating cycle. The inset shows DSC thermograms for DPPC ULVs in the absence and presence of HMIM[Br] and DMIM[Br] in the heating cycle.

The effectiveness of ILs in permeabilizing the DPPC ULVs greatly relies on the extent of the IL-membrane interaction. To assess the strength of this interaction, isothermal titration calorimetry (ITC) measurements were carried out. ITC titration curves or thermograms for the interaction of DMIM[Br] and HMIM[Br] with DPPC membrane are shown in Fig. S1. Binding isotherms are constructed from the ITC titration curves by integrating over all the peaks in the thermograms (Fig. S1). These isotherms are plotted as a function of IL/DPPC molar ratio in Fig. 3(b) for both HMIM[Br] and DMIM[Br]. It is evident that in contrast to DMIM[Br], titration of HMIM[Br] showed insignificant molar heat values, suggesting a relatively weaker interaction with the lipid membrane. While both ILs exhibit similar electrostatic interactions with the lipids, their distinct binding strengths are attributed to difference in hydrophobic interactions arising from varying chain lengths. This highlights the critical role of alkyl chain length in IL-lipid membrane interactions, with longer chains enhancing the interaction strength, consistent with the dye-release assay measurements. The ITC data for DMIM[Br] is analyzed by fitting the isotherms using a one-site binding model[26] which assumes the presence of independent binding sites for ILs. Details of ITC data analysis are given in supporting information (SI). The association constant ($K$) for DMIM[Br]-DPPC membrane interaction at 320 K is found to be $9.4\times 10^3$ M$^{-1}$ and the Gibbs free energy associated to binding of DMIM[Br] with the DPPC membrane is calculated, using $\Delta G = -RT \ln 55.5K$, to be -8.4 kcal/mol.

DPPC membranes exist into two major phases, an ordered gel phase below 315 K and a fluid phase above this temperature, more relevant for physiological conditions[27]. ITC measurements were also carried out with DMIM[Br] in ordered gel phase (at 300 K) of DPPC and observed thermograms are shown in the inset of Fig 3 (b). Surprisingly, an almost negligible interaction is observed in the gel phase compared to the fluid phase, suggesting an important dependence of the IL-membrane interactions on the membrane's physical state. To gain further microscopic insights, molecular dynamics (MD) simulations were performed on DPPC bilayers with DMIM[Br] in both gel (300 K) and fluid (320 K) phases. Initially, ILs were solvated within 5 Å proximity to both upper and lower leaflets of the bilayer, evenly distributed above and below. Snapshots of equilibrated IL-membrane systems (as shown in Fig. 3(c)) elucidated that in the fluid phase, DMIM[Br] readily penetrates the lipid membrane, positioning its alkyl chains within the membrane core and the imidazolium ring near the lipid head-water interface. In the gel phase, which is characterized by highly ordered and tightly packed DPPC molecules, the insertion of the IL within the membrane is clearly hindered. As membrane insertion incurs higher energy costs than into the water layer, most IL molecules fail to penetrate the lipid bilayer. This is consistent with our assessment from dye leakage measurements which show almost no leakage in the case of DMIM[Br] for DPPC ULVs at 300 K. MD simulations results are also consistent with ITC data, which indicated no significant binding of DMIM[Br] in the gel phase. Furthermore, our MD simulations revealed a significant observation: once the ILs are fully inserted in the fluid phase, a reduction in temperature to the gel phase does not lead to the expulsion of IL molecules from the membrane. This suggests that the initial phase of the membrane is pivotal for IL penetration during its solvation.

Incorporation of ILs into the lipid membrane is likely to affect the structure and phase behavior of the vesicles[16, 18, 28-30]. Dynamic light scattering (DLS) measurements reveal that the addition of ILs affects the size and polydispersity of the vesicles (See Fig. S2 in supporting information), with stronger effects in the case of the longer chain IL, DMIM[Br]. Differential scanning calorimetry (DSC) thermograms for DPPC vesicles in the absence and presence of HMIM[Br] and DMIM[Br] are shown in the inset of Fig. 3(d). As expected DPPC shows a distinct, sharp peak (FWHM=0.8 K) at 315 K, corresponding to the main phase transition ($T_m$). Addition of IL leads to a shift downwards in the main transition temperature, to 314 K and 311 K for HMIM[Br] and DMIM[Br] respectively, and a broadening of the peaks (FWHM=1.3 K and 2.6 K for HMIM[Br] and DMIM[Br] respectively). As peak width is inversely proportional to the transition cooperativity, the results suggest that the IL reduces the cooperativity of the main phase transition, more so in the case of longer chain IL, DMIM[Br]. Moreover, the peak corresponding to the main phase transition in the presence of DMIM[Br] is found to be slightly asymmetric which could indicate the formation of phase separated domains corresponding to IL rich and IL poor phases, as directly observed using small angle X-ray diffraction[31-32] and [31]P-NMR[33]. These domains have very close melting temperatures hence can lead to an asymmetry in the transition peak. These domains of compact arrangement of lipids may cause low areal number density of lipid in the remaining part of the membrane. It may cause overall enhancement of the membrane permeability. It has been shown that IL induced phase segregation leads to formation of interdigitated gel phase in the matrix of membrane of zwitterionic lipids[21,31-32]. In addition, the pre-transition is weakened for the HMIM[Br] system and is no longer observed with DMIM[Br]. Despite the observed shift in $T_m$ towards lower temperatures and peak broadening, the addition of ILs does not significantly influence the enthalpy ($\Delta H$), which is found to be the same (~ 31 kJ/mol) for pristine DPPC and DPPC in the presence of ILs. As $T_m$ decreases for DPPC with ILs, according to $\Delta H = T_m \Delta S$, the presence of ILs reflects an increase in the entropy[34]. The effect is particularly pronounced for the longer alkyl chain IL, DMIM[Br] for which $T_m$ decreases from 315 K to 311 K which leads to increase in $\Delta S$ from 98 J/mol/K to 100 J/mol/K.

DSC provides thermodynamic information and indirectly informs of structural changes with temperature, however, it does not provide any direct information about changes in the dynamics of the lipids. For this, elastic fixed window scans (EFWS) as measured using neutron spectroscopy have been carried out on DPPC membranes in the absence and presence of DMIM[Br] and the data are shown in Fig. 3 (d). In an EFWS, the neutron elastic intensity at near zero energy transfer, determined by the spectrometer's energy resolution, is monitored as a function of temperature. As the temperature increases and molecules become more mobile, the elastic intensity decreases. Any abrupt or sharp changes in the elastic intensity indicate significant alterations in dynamics, often associated with phase changes. As seen in the figure, the elastic intensity for pure DPPC decreases monotonously as the temperature is increased, until ~ 315 K where there is a sudden drop. This temperature corresponds to the $T_m$ and suggests that the main phase transition involves a change in lipid molecule dynamics as well as a structural rearrangement. Similarly, for DPPC with DMIM[Br], a sharp drop is observed at a temperature of ~ 310 K consistent with the findings from DSC. EFWS measurements indicate that the incorporation of IL into the membrane has a discernible impact on its dynamics. Quasielastic neutron scattering (QENS) can be used to gain further microscopic insights into the underlying dynamical changes. QENS gives access to dynamics in the picoseconds to nanoseconds time scale and at length scales ranging from angstroms to tens of nanometers[27, 35-40]. To this end we performed QENS measurements on DPPC membranes with and without DMIM[Br] and HMIM[Br], in the fluid phase at 330 K. Typical observed QENS data for DPPC vesicle solution and solvent ($D_2O$) are shown in Fig. S3. The spectra for the DPPC membrane can be obtained by subtracting the solvent contribution from the vesicle solution. Typical QENS spectra for the DPPC membrane in the absence and presence of 20 wt % DMIM[Br] are shown in Figure 4 (a). The instrument resolution as measured with a vanadium standard is shown by the dashed line. For direct comparison, spectra are normalized to the peak amplitudes. The DPPC membrane presents significant dynamical activity in the spatio-temporal range of the IRIS spectrometer[41], as revealed by the strong quasielastic broadening, with the incorporation of the IL accelerating the membrane dynamics. Within the spatio-temporal domain of IRIS spectrometer, lipid molecules predominantly exhibit two distinct dynamical modes: (i) lateral motion within the leaflet and (ii) localized internal motions of lipid alkyl chains[27, 37, 42-45]. The observed scattering data can be described as a convolution of two scattering laws, each associated with one of the two motions , expressed as[46]:

$$S_{mem}(Q,E) = A(Q)L_{lat}(\Gamma_{lat}, E) + (1 - A(Q)L_{tot}(\Gamma_{lat} + \Gamma_{int}, E)$$

(1)

Here $A(Q)$ represents the elastic incoherent structure factor (EISF) of the internal motion which gives information about the geometry of the internal motions, and $\Gamma_{lat}$ and $\Gamma_{int}$ are the half-width-at-half-maximum (HWHM) of each Lorentzian functions, which informs on the time scale of these motions. Equation (1) was convoluted with the instrument resolution function, and the values of $A(Q)$, $\Gamma_{lat}$, and $\Gamma_{int}$ were determined through least-square fitting of the QENS spectra. A representative fit is shown for DPPC membrane with 10 wt % DMIM[Br] in Figure 4 (b), distinguishing the contributions from the two dynamical modes. Obtained EISF and HWHM corresponding to internal motion of DPPC membrane in the absence and presence of ILs are presented in Fig. S4. Internal motion of lipid is analyzed assuming localized translational diffusion and obtained results are discussed in SI.

The lateral diffusion of lipids holds pivotal importance in various physiologically relevant membrane functions, including cell signaling and membrane trafficking. In addition, lateral motion plays a crucial role in regulating the fluidity of the cell membrane, which in turn governs its permeability. As a semipermeable barrier, the cell membrane selectively allows essential elements to travel in and out of cells while blocking harmful components from entering the cells. Thus the cell membrane needs to maintain a delicate balance between fluidity, which enables proteins and lipids to move as required to perform their functionalities, without compromising membrane integrity or allowing substances to leak in or out of the cell. Even subtle alterations in lipid lateral diffusion can significantly impact both the fluidity and the permeability of the cell membrane, potentially influencing its transport properties and functionalities. These changes ultimately dictate the stability of bacterial cells. Restricted lateral diffusion of lipids can decrease membrane fluidity, potentially reducing membrane permeability and impeding the transport of vital components. Conversely, accelerated lateral diffusion can enhance membrane fluidity, thereby increasing membrane permeability and potentially allowing harmful substances to infiltrate cells or leakage of intracellular cytoplasm, leading to cellular damage. Therefore, the functionality of cell membranes is intricately linked to the lateral diffusion of lipids. The HWHMs of the Lorentzian associated to lateral motion for DPPC membrane in the absence and presence of two different ILs are shown in Fig. 4 (c). Incorporation of IL into the membrane results in increased $\Gamma_{lat}$ values, suggesting faster lateral motion, which further increases with increasing IL concentration. At the same concentration, addition of a shorter alkyl chain length IL, HMIM[Br], accelerates the lateral motion to a lesser extent than DMIM[Br]. For all the systems, $\Gamma_{lat}$ exhibits a linear increase with $Q^2$ passing through the origin, indicative of continuous diffusion, as described by Fick's law, $\Gamma_{lat} = D_{lat}Q^2$. The lateral diffusion coefficients, $D_{lat}$, obtained from the slope of these linear fits, are presented in Figure 4(d). The addition of both ILs clearly enhances the lateral diffusion coefficients of the lipids.

The cell membrane in any native organism has an optimum permeability. Any increase in permeability compromises the structural and functional integrity of any cell, including bacterial cells, which can lead to disruption of transmembrane electrochemical gradients and osmotic balance, leakage of cytoplasm, influx of toxic substances, inhibition of energy production, and/or impairment of essential cellular processes. These cumulative effects will culminate in bacterial cell death, highlighting the significance of membrane permeability or lateral diffusion of lipid. This lateral diffusion of lipids within a leaflet is dependent on several factors including the area per lipid, the membrane curvature, the surface charge density of the membrane, the interaction between additive and lipids, and obstructions [35, 37, 44-46]. X-ray/neutron reflectivity measurements reveal that incorporating ILs into the membrane results in a reduction in the thickness of the membrane bilayer [23,47], also corroborated by MD simulations[47]. As the volume compressibility of lipids is insignificant[48-49], the decrease in membrane thickness, induces an increase in area per lipid. According to free volume theory, lateral diffusion is limited by the occurrence of a free volume greater than a critical size next to a diffusing particle[50-51]. It suggests that lateral diffusion of lipids strongly depends on the area per lipid, increasing with increasing area per lipid.

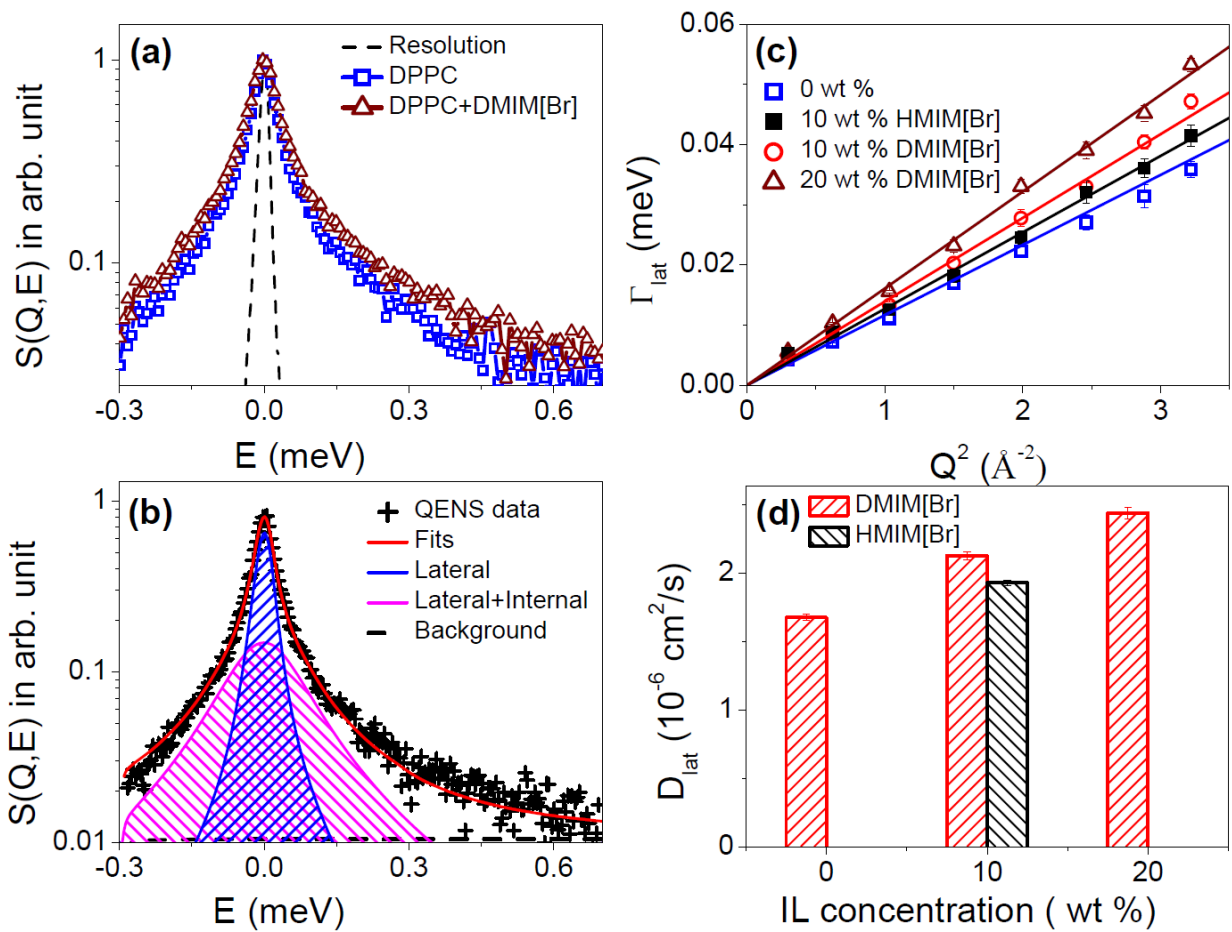

Fig. 4 (a) Representative QENS Spectra at a momentum transfer, $Q$, of 1.2 Å⁻¹ for DPPC in the absence and presence of 20 wt % DMIM[Br] in the fluid phase (at 330 K). The contribution of the solvent ($D_2O$) has been subtracted from the data. The instrument resolution as measured with a vanadium standard is shown by the dashed line. For quantitative comparison, the spectra are normalised to peak amplitude. (b) Individual fit contributions shown for the QENS spectra at $Q$ = 1.0 Å⁻¹ for DPPC with 10 wt % DMIM[Br] at 330 K described by Eq. (1) . (c) Variation of the half width at half maximum (HWHM) of the Lorentzian representing lateral lipid motion ($\Gamma_{lat}$) with respect to $Q^2$ for DPPC membranes in the absence and presence of ILs. The lines represent fits with the Fickian diffusion model discussed in the text. (d) Lateral diffusion coefficient, $D_{lat}$, for DPPC membrane at varying concentrations of DMIM[Br] IL in the fluid phase (330 K). $D_{lat}$ obtained for DPPC membrane with 10 wt % HMIM[Br] is also shown for direct comparison.

Enhancement of lateral motion depends on the chain length of the IL; incorporation of 10 wt % HMIM[Br] increases $D_{lat}$ by 15 % ( from $1.68 \pm 0.02 \times 10^{-6}$ cm$^2$/s to $1.93 \pm 0.02 \times 10^{-6}$ cm$^2$/s) compared to the addition of DMIM[Br], which leads to an increase of $D_{lat}$ by 27 % ( from $1.68 \pm 0.02 \times 10^{-6}$ cm$^2$/s to $2.13 \pm 0.02 \times 10^{-6}$ cm$^2$/s). DMIM[Br], as the longer chain IL, has a stronger binding affinity to the lipid membrane as shown by our ITC measurements, causing greater disruption to the lipid arrangements. This also correlates well with the higher permeability in dye leakage assay and antimicrobial activity shown by DMIM[Br]. Increasing the concentration of IL in the lipid further accelerates lateral motion. At a concentration of 20 wt % DMIM[Br], $D_{lat}$ undergoes a significant 45 % increase, rising from $1.68 \pm 0.02 \times 10^{-6}$ cm$^2$/s to $2.44 \pm 0.04 \times 10^{-6}$ cm$^2$/s. This aligns with the observed trend of membrane thinning with increasing IL concentration[23,47], resulting in an expanded area per lipid and consequently facilitating lateral diffusion. These observations compellingly underscore the robust and intricate correlation between the structural characteristics and dynamic behaviors within these systems, elucidating the profound interplay between the arrangements of lipids and fluidity of the membranes.

In conclusion, our comprehensive investigation has revealed the diverse antimicrobial potential of ILs, intimately tied to their interactions with lipid membranes. The observed decrease in MIC with an increase in the alkyl chain length of ILs highlights the significance of hydrophobic interactions with the lipid membrane as the predominant driving force, a conclusion further supported by binding affinities observed through ITC. Flow cytometry and dye leakage experiments have elucidated that ILs directly interact with the lipid membrane, resulting in cell permeabilization and suggesting that as the primary mechanism underlying their bactericidal action. Our findings provide the rationale behind the antimicrobial activity of ILs over a wide range of pathogens, as main action mechanism directly involves lipid membrane. Encouragingly, haemolysis measurements demonstrate a lack of toxicity even at concentrations up to twice of MBC, underscoring the selective action of ILs towards bacterial membranes. Furthermore, QENS measurements have revealed that the incorporation of ILs enhances the lateral diffusion of lipids, a pivotal factor contributing to enhanced fluidity and membrane permeability. The more pronounced effects of ILs on lateral diffusion highlight a stronger IL-membrane interaction, particularly with longer alkyl chain ILs. This observed pattern aligns seamlessly with binding affinity, membrane phase behavior, membrane permeability, and antimicrobial activity, reinforcing the interconnectedness between them. This study offers a hierarchical overview of antimicrobial activity of ILs by providing explanations rooted in microscopic physics and linking them to their biological action mechanism. In essence, our findings unravel physical insights into the bactericidal action mechanism of ILs, offering promise for advancing antimicrobial strategies through the rational design and optimization of IL-based antimicrobial agents to effectively combat resistant pathogens in healthcare applications.

**MATERIALS AND METHODS**

The phospholipid DPPC was purchased from Avanti Polar Lipids (Alabaster, AL) in powder form. Ionic liquids EMIM[Br], HMIM[Br] , DMIM[Br] and DMIM[BF4] were obtained from TCI Chemical. Propidium iodide, $D_2O$, dialysis tubing, and calcein were procured from Sigma Aldrich. Resazurin was purchased from Himedia (Mumbai, India).

MIC and MBC of ILs were determined using resazurin reduction assay and spot assay, respectively. To evaluate the toxicity of ILs towards human cells and understand their action mechanism, hemolysis and flow cytometry measurements were conducted. Further details of these experiments can be found in the SI.

DPPC ULVs in the absence and presence of ILs were prepared using the extrusion method following the protocol described in the SI. Dye leakage assay measurements were carried out using a green fluorescent dye calcein. The dye release with time was monitored by following the change in fluorescence intensity upon interaction of IL with the lipid membrane using a plate reader (Polarstar omega, BMG Labtech, Offenburg, Germany). To investigate the thermodynamics of binding of the ILs to the DPPC ULVs, ITC measurements were carried out using a MicroCal iTC200 system from Malvern Instruments. Details of the dye-leakage assay and the ITC experiments are given in SI.

To investigate the effects of the ILs on the size, stability and phase behaviour of vesicles, DLS and DSC measurements were carried out on DPPC ULVs in the absence and presence of ILs. DLS and DSC measurements were performed using a Zetasizer Nano ZS system (Malvern Instruments, UK) and Mettler Toledo DSC instrument, respectively. Comprehensive information regarding the DLS and DSC experiments is given in SI.

Elastic incoherent neutron scattering (EINS) and QENS experiments were carried out on DPPC ULVs with and without ILs using the inverted geometry spectrometer IRIS[41] at the ISIS pulsed Neutron and Muon source at the Rutherford Appleton Laboratory, UK. Details of the EINS and QENS experiments can be found from SI.

MD simulations were conducted using the NAMD simulation package[52] on a DPPC lipid bilayer comprising 128 lipids (64 per leaflet) and solvated with a water-to-lipid ratio of 111:1, both in the absence and presence of ILs. Further details regarding the MD simulations are given in the SI.

**SUPPORTING INFORMATION (SI)**

Additional details of the experiments, simulations and data analysis are given in SI. ITC titration curves for IL-DPPC interactions, DLS data on DPPC ULVs in absence and presence of ILs, QENS spectra for DPPC vesicles solution and solvent are shown in Fig S1, S2 and S3, respectively. EISF and HWHM corresponding to internal motion of DPPC lipid in absence and presence of IL are shown in Fig. S4.